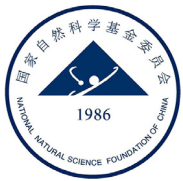

Contents lists available at ScienceDirect

# Fundamental Research

journal homepage: http://www.keaipublishing.com/en/journals/fundamental-research/

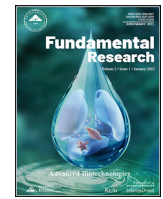

Article

# Unexpected type-II multiferroic phase in GdMnO$_3$ under high magnetic fields

Ming Yang [a,1], Jun Chen [b,1], Junfeng Wang [a,*], Chao Dong [a], Chengliang Lu [a], Gang Xu [a], Jinguang Cheng [c], Jianshi Zhou [d], Shuai Dong [b,*]

[a] *Wuhan National High Magnetic Field Center and School of Physics, Huazhong University of Science and Technology, Wuhan 430074, China*
[b] *Key Laboratory of Quantum Materials and Devices of Ministry of Education, School of Physics, Southeast University, Nanjing 211189, China*
[c] *Beijing National Laboratory for Condensed Matter Physics and Institute of Physics, Chinese Academy of Sciences, Beijing 100190, China*
[d] *Materials Science and Engineering Program, Mechanical Engineering, University of Texas at Austin, Texas 78712, United States*

ABSTRACT

Perovskite manganites with small A-site ions, as the first and canonical branch of type-II multiferroics, are ideal systems to exhibit magnetism-induced ferroelectricity. Despite their established magnetoelectric phase diagrams under low magnetic fields, here an unidentified phase with a large magnetism-induced polarization (up to 1500 μC/m$^2$) is revealed in GdMnO$_3$ under high magnetic fields up to 60 T. Based on multiprobe experiments, a complete phase diagram is constructed with successive polar-nonpolar-polar-nonpolar transitions. Such a non-monotonic evolution is well mimicked by model simulation, while the spin-lattice coupling is the key ingredient for the reentrant ferroelectric phase.

## 1. Introduction

Orthorhombic rare-earth manganites, with general formulas $R$MnO$_3$ and $R$Mn$_2$O$_5$, set off an intense period of interest in type II multiferroics [1,2], which co-establish a fertile playground to explore the multiferroicity and colossal magnetoelectricity [3–6]. As canonical type-II multiferroics, their special physical properties have been systematically revealed, and more importantly their magnetoelectric mechanisms also shed light on the whole field of multiferroics [7–10].

Although tremendous interests have been drawn to investigate the complex magnetic orders and related ferroelectricity in these systems [11–15], unexpected magnetoelectric states/behaviors with new physics are continually emerging. For example, very recently, GdMn$_2$O$_5$ was reported to exhibit a topologically protected switching phenomenon, that a four-state polarization flop is realized at a magic angle under strong magnetic fields [16,17].

In contrast to the complicated $R$Mn$_2$O$_5$, $R$MnO$_3$'s are relative simpler regarding its crystalline and magnetic structures, which have been generally understood as follows. For those relative large $R$ ions, e.g. La$^{3+}$ and Pr$^{3+}$, the ground state of $R$MnO$_3$ is A-type antiferromagnetic (A-AFM), which is not ferroelectric [18,19]. Weak ferromagnetism also occurs from the spin canting. For those middle-size $R^{3+}$'s, e.g. Tb$^{3+}$ and Dy$^{3+}$, the magnetic ground states become the cycloid spiral (SP) ones [11,12,19], which induce polarizations (in the order of ∼ 1000 μC/m$^2$) via the inverse Dzyaloshinskii-Moriya interaction (DMI) [20,21]. For those small-size $R^{3+}$'s, e.g. Ho$^{3+}$ and Lu$^{3+}$, the magnetic order at low temperatures is E-type antiferromagnetic (E-AFM), which can generate large polarizations in the order of ∼ 5000 − 10000 μC/m$^2$ via the exchange striction mechanism [22,23]. Such a phase evolution as a function of $R^{3+}$ size can be well reproduced by the double-exchange model or classical spin model [24–27], both of which rely on the subtle competition between the nearest-neighbor and next-nearest-neighbor interactions.

For those compounds near phase boundaries, the magnetoelectric behaviors can be nontrivial. For example, for $R$ = Eu, there is a field-induced ferroelectric SP phase under high magnetic fields 22 T< $H$ < 50 T, although its zero-field ground state is paraelectric and A-AFM (and weakly ferromagnetic) [28]. Similar behavior was also reported for GdMnO$_3$ with a much lower threshold of magnetic field (∼ 1 T), since it is very proximate to the boundary between the A-AFM and SP phases

---

*  Corresponding authors.
   *E-mail addresses:* jfwang@hust.edu.cn (J. Wang), sdong@seu.edu.cn (S. Dong).
[1] These authors contributed equally to this work.

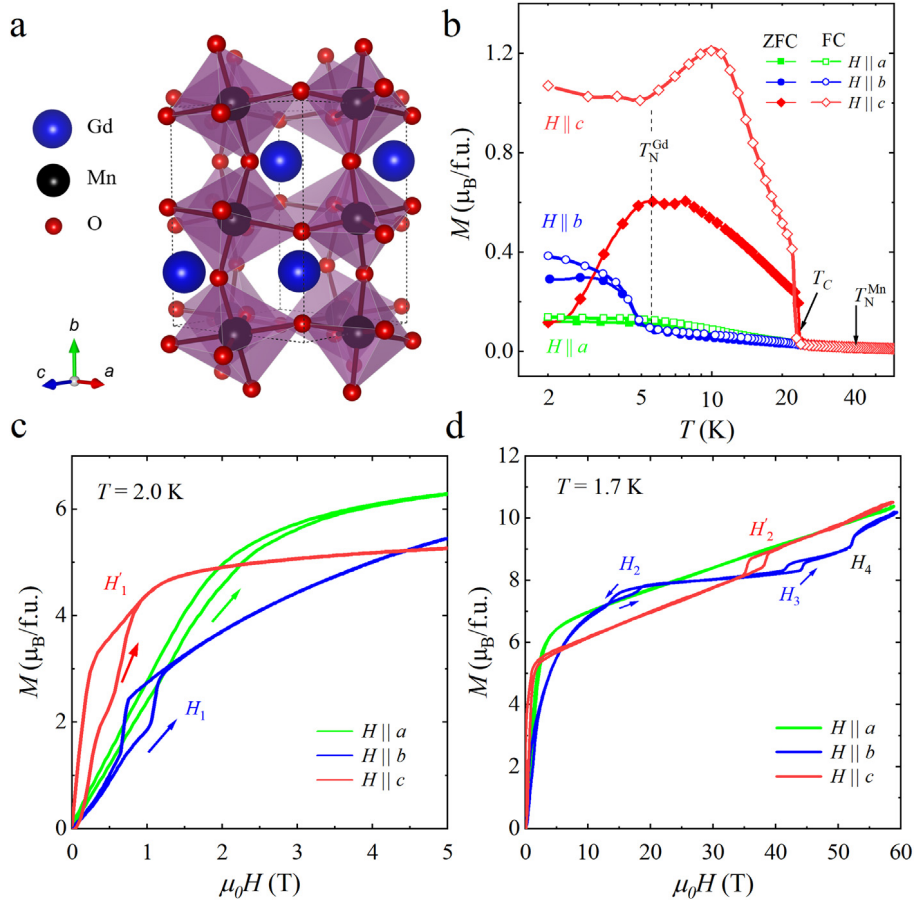

**Fig. 1.** Magnetization $M$ for magnetic field along different crystalline axes at 1.7 K. The inset shows $M$ at low field range which are measured by SQUID at 2 K. The arrows indicate the sweeping direction of magnetic fields.

[12]. Then is the magnetoelectric story of $R$MnO$_3$'s fully completed after their history of two decades?

In this work, a systematic high-field study is performed on GdMnO$_3$ single crystals. With pulsed fields up to 60 T, a hidden magnetoelectric phase is uncovered in the high-field region. A novel polar-nonpolar-polar-nonpolar transition sequence occurs with increasing magnetic field. And the phase diagram can be mimicked by a spin model with multiple interactions, which predicts an orthogonal spin stripy phase as the second polar phase. The spin-lattice coupling is the key ingredient for the multiferroicity of new phase.

## 2. Methods

Single crystals were grown by the floating-zone method as described elsewhere [29]. Crystals were orientated by the Laue X-ray photography and cut into cubes with dimensions 2 mm × 2 mm × 2 mm for magnetization measurements and plates 2 mm × 2 mm × 0.2 mm for electric polarization and magnetostriction measurements, respectively. High magnetic field measurements were performed in a 12 ms-pulsed magnet at Wuhan National High Magnetic Field Center. The magnetization was measured by the induction method using a coaxial pickup coil and calibrated by the low-field data measured by a superconducting quantum interference device (Quantum Design SQUID). The electric polarization was probed by measuring and integrating the pyroelectric current flowing through the two largest surfaces of a platelike sample with electrodes attached by sliver paste [30,31]. During the polarization measurements, a bias voltage of 100 V, which converts into a bias field of $E = +500$ kV/m, was applied to polarize the FE domains induced by external magnetic fields. Magnetostriction along the field direction was measured by the fiber-Bragg-grating (FBG) method [32].

## 3. Experimental results

Fig. 1 shows the magnetization ($M$) along the crystallographic $a$, $b$, and $c$ axes of GdMnO$_3$ at 1.7 K. The low field results, as shown in the inset of Fig. 1 and Fig. S1 in Supplementary Materials (SM) [33], are consistent with previous reports [12,34], which indicate the $c$-axis as the macroscopic easy magnetization axis. Since the ground state of GdMnO$_3$ is antiferromagnetic, the microscopic easy axis of spins should be perpendicular to the easy magnetization axis, i.e., lying in the $ab$ plane. Remarkable hysteresis behaviors between the increasing and decreasing fields are discernible under 3 T along all directions, indicating a first-order transition in the low-field region. As reported before, this transition was attributed to the A-AFM to SP phase transition.

For $H||a$, $M$ steeply increases up to ~ 6 $\mu_B$ per formula unit (f.u.) at 3 T and then rises linearly without any anomaly nor saturation till the maximum field. Due to the very localized electronic cloud of $4f$ orbitals, the nearest neighboring exchanges between $4f - 4f$ spins (as well as between $4f - 3d$ spins) are naturally weak (usually < 1 meV). Thus the magnetic moments of Gd$^{3+}$ are much easier to be aligned ferromagnetically by external magnetic field. Noting that the saturated moments of Gd$^{3+}$ and Mn$^{3+}$ are 7 $\mu_B$ and 4 $\mu_B$, respectively. Then the low-field steep increase of magnetization can only be mainly attributed to the parallel alignment of Gd$^{3+}$ spins. The slightly reduced magnetization from the saturation value (7 $\mu_B$ per f.u.) is due to the antiparellel contribution from Mn$^{3+}$'s canting spins [34].

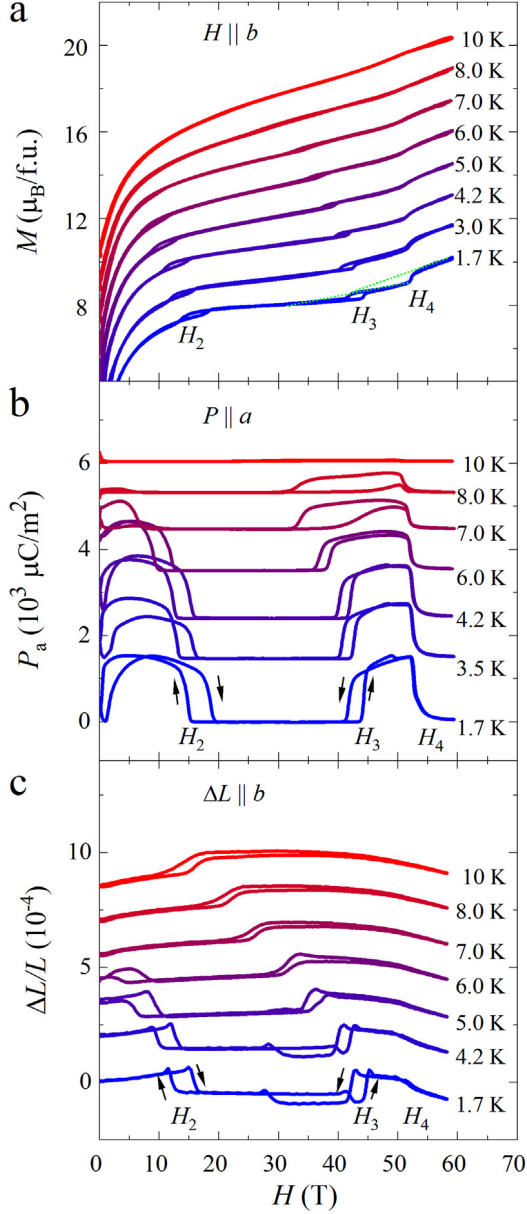

**Fig. 2.** (a) Magnetization $M$, (b) polarization $P_a$, and (c) magnetostriction $\Delta L/L$ along the $b$-axis at various temperatures for $H||b$ up to 60 T. All curves are vertically shifted for clarity.

In contrast to the featureless magnetization vs field curves of $H||a$, a series of successive magnetic transitions are found at $H_1 \sim 0.6$ T, $H_2 \sim 14$ T, $H_3 \sim 41$ T, and $H_4 \sim 52$ T for $H||b$. The first three transitions at $H_1$, $H_2$, and $H_3$ are all in the first-order, characterized by the presence of hysteresis loops. However, the hysteresis at $H_4$ becomes immeasurably small and thus this transition is likely second order at the measured temperatures. Interestingly, an approximative plateau of magnetization is observed between $H_2$ and $H_3$, indicating a gaped state in this range.

For comparison, when $H$ is applied along the $c$ axis, only another spin-flip transition can be found at $H'_2 \sim 35$ T, in addition to the first one at $H'_1 \sim 0.8$ T. Furthermore, the two $M-H$ curves of the $H||a$ and $H||c$ cases coincide with each other when $H > H'_2$, implying the magnetic isotropy in the $ac$ plane above this field. Another conclusion is that a full saturation magnetization (11 $\mu_B$/f.u.) has not been reached till 60 T, although it is indeed approached (> 10 $\mu_B$/f.u.). As stated before, since $Gd^{3+}$'s spins are almost fully aligned ferromagnetically above 3 T,

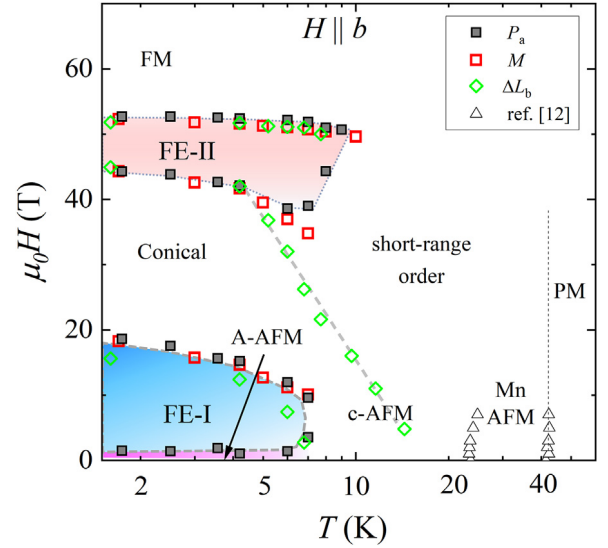

**Fig. 3.** The magnetoelectric phase diagram of GdMnO$_3$. Circles and squares denote the transition fields on the $M(H)$ and $P(H)$ curves. Open and filled symbols represent the increasing and decreasing field, respectively. Open triangles are extracted from ref. [12].

these high-field transitions should be attributed to the re-ordering of the $Mn^{3+}$'s spins.

The successive anomalies in the $b$-axis magnetization indicate the unknown magnetic transitions under high fields, which are beyond all previous reports and certainly deserve more investigations. The temperature dependence of $M-H$ curves are plotted in Fig. 2a. First, with increasing temperature, all anomalies become weaker and weaker and finally disappear at around 8 K. Second, both the transitions at $H_2$ and $H_3$ move towards lower fields, while the transition at $H_4$ persists without obvious shifting.

To reveal more information of these high-field transitions of GdMnO$_3$, the electric polarization along the $a$-axis ($P_a$) is measured as a function of $H||b$, as shown in Fig. 2b. It is clear that a non-monotonic evolution of $P_a$ with multiple steps coincides well with the $M$ curves [Fig. 2a]. At $T = 1.7$ K, a finite $P_a$ emerges rapidly since $H_1$, consistent with previous reports [12,35], which was successfully ascribed to the field-induced $ab$-plane cycloidal structure of $Mn^{3+}$ spins. Unexpectedly, this $P_a$ decreases to zero at $H_2$ and the system restores to the paraelectric state till $H_3$. Even more surprisingly, the $P_a$ emerges again in the high-field region between $H_3$ and $H_4$, and the paraelectric state restores again beyond $H_4$. Such a nontrivial evolution defines five phases. As aforementioned, the first two below $H_2$ had been understood and the last one beyond $H_4$ is almost ferromagnetic. Then the rest two between $H_2$ and $H_4$ should be clarified.

The magnitude of high-field $P_a$ is comparable to the low-field one, reaching $\sim 1500$ μC/m$^2$. Hystereses of $P_a$ exist at $H_1$, $H_2$, and $H_3$, while non-hysteresis occurs at $H_4$. These behaviors, as well as the their temperature-dependences, coincide with aforementioned $M-H$ curves.

The magnetostriction $\Delta L/L$ along $b$-axis is also characterized. As shown in Fig. 2c, at $T = 1.7$ K, the transition fields, plateaus, and hystereses of $\Delta L/L$ generally match well with the behaviors of $M$ and $P_a$. Another interesting behavior is that the hystereses of $\Delta L/L$ are broader, which even persist at relative high temperatures (e.g. 10 K). These differences between magnetostriction and polarization may come from the complex domain dynamics during those first-order transitions.

In summary of experimental results, the spin re-ordering under high fields are responsible for the non-monotonic evolution of $P_a$. Based on above results, a magnetoelectric phase diagram of GdMnO$_3$ is sketched in Fig. 3, while the phase boundaries are defined by aforementioned transitions.

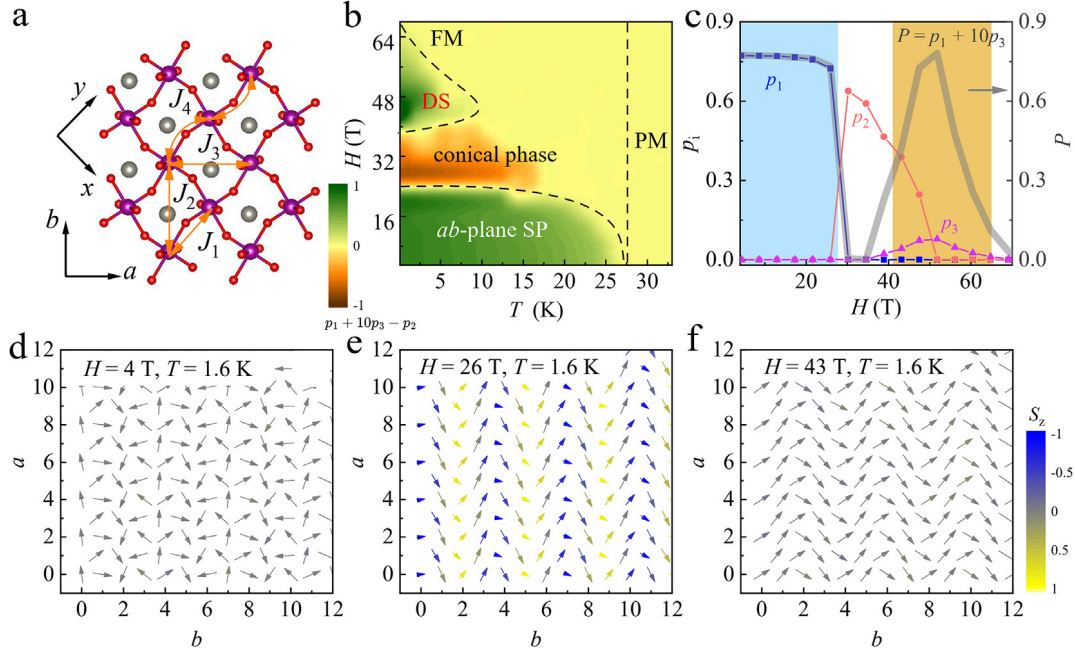

**Fig. 4.** (a) Top view of GdMnO$_3$ in *ab*-plane and labelled exchange interaction $J_1$, $J_2$, $J_3$ and $J_4$. (b) Phase diagram in $T - H$ space with color contour map determined by the value $p_1 + 10p_3 - p_2$; (c) The evolution of order parameters $p_1 = \langle|\Sigma_{i,\hat{u}=\hat{x},\hat{y}}\mathbf{S}_i \cdot (\mathbf{S}_{i-\hat{u}} - \mathbf{S}_{i+\hat{u}})|\rangle/4N$, $p_2 = \langle|\Sigma_{i,\hat{u}=\hat{x},\hat{y}}\hat{u} \cdot (\mathbf{S}_i \times \mathbf{S}_{i+\hat{u}})|\rangle/2N$, $p_3 = \langle|\Sigma_{i,\hat{u}=\hat{x},\hat{y}}\hat{z} \cdot (\mathbf{S}_i \times \mathbf{S}_{i+\hat{u}})|\rangle/2N$ and the weighted parameter $P$ at low temperature (1.6 K) as functions of magnetic field. (c-e) The snapshots of one-layer spin textures in the region. (c) The *ab*-plane spiral phase (SP). (d) The conical phase. (e) The orthogonal antiferromagnetic double stripes (DS) phase. Only in the conical phase, the out-of-plane spin components are nonzero, as indicated by the color contour map.

## 4. Model simulations and discussion

Although the reentrances of multiferroic phases under high magnetic fields have already been observed in other frustrated magnetic systems like Ni$_3$V$_2$O$_8$, $A_2$V$_2$O$_7$ ($A$=Ni, Co), and CuCrO$_2$ [30,36–39], their underlying mechanisms seem to be unambiguously different. The magnitude of these high-field polarization is in the order of 100 μC/m$^2$ or even smaller, which should be orignated from the inverse DMI and field-induced SP phases. In other words, the reentrant multiferroic phases in those systems are more similar to the first multiferroic phase in GdMnO$_3$. Therefore, the high-field multiferroic phase in GdMnO$_3$, with a much larger polarization up to 1500 μC/m$^2$, should be understood with an alternative scenario.

In the current stage, the common techiques to directly characterize magnetic orders, like neutron scattering, are not available under such high magnetic fields. Thus, a model simulation becomes the preferred approach to access more details of the high-field phase. In fact, Mochizuki et al. developed a classical spin model for $R$MnO$_3$'s [25,26,40,41], which can precisely describe their complex phase diagram in the low-field region. Based on Mochizuki et al.'s model, a simplified model Hamiltonian is used here:

$$H = \sum_{i,j} J_n \mathbf{S}_i \cdot \mathbf{S}_j - J_b \sum_{i,j} (\mathbf{S}_i \cdot \mathbf{S}_j)^2 + \sum_i \left[ A_c (S_i^z)^2 - A_\delta (S_i^z)^4 \right]$$
$$- J_\delta \sum_i \sum_{u=x,y} (\mathbf{S}_i \cdot (\mathbf{S}_{i+u} - \mathbf{S}_{i-u}))^2 + \mathbf{H} \cdot \sum_i \mathbf{S}_i, \tag{1}$$

where the first is the neighboring exchanges between Mn's spins (**S**'s), including in-plane $J_1/J_2/J_3/J_4$ [see Fig. S5(a)] and out-of-plane $J_c$. The second term is biquadratic interaction merely for the nearest neighboring spins. The third term includes conventional single-ion anisotropy and cubic anisotropy. The fourth term is an effective expression for spin-lattice coupling (i.e. the exchange striction), which prefers the E-AFM order. The last one is Zeeman energy under magnetic field **H** along *b* direction. In our model, only Mn's sublattice is considered, thus the low-field magnetic phase (below 2 T) could not be interpreted while Gd's sublattice is only a ferromagnetic background.

We apply first-principles calculations to determine the realistic relationship among all the exchange interactions, then with proper coefficients choice and using Monte Carlo simulations [more details refer to SM [33]], the phase diagram can be well reproduced (even not every detail), as shown in Fig. 4b. The phase regions are determined by the order parameters, as shown in Fig. 4c. According to our simulations, at low temperatures and in the low-field region, the magnetic state is belonging to the *ab*-plane spiral phase with incommensurate modulation vector $q \approx 0.27$ as performed in Fig. 4d, which almost matches the previous experimentally measured $q = 1/4$ [12]. With increasing magnetic field, the system turns to be the conical phase [Fig. 4e]. Further increasing field, the spin texture forms special orthogonal antiferromagnetic double stripes (DS), as characterized as "↑↑↑↑→→→→" lying in the *ab* plane [Fig. 4f], which generally breaks the inversion symmetry to become a novel promising FE phase. In fact, the DS order has already been suggested as a possible ground state in the earlier study[12] and also similar to a previously predicted "spin-orthogonal stripe" although not identical [42]. It's noteworthy that the presence of DS phase is significantly dependent on spin-lattice coupling with a certain strength [see Fig.S6b], rendering a subtle competition of magnetic phases. In addition, the mentioned phase transitions simultaneously contribute to the sudden change of magnetization as illustrated in Fig. S7, aligning with experiments well. For all these three orders, their modulation vectors are along the *b*-axis, and the interlayer coupling also changes from antiferromagnetic and ferromagnetic gradually. In the very high-field limit, the system finally becomes ferromagnetic.

The associated polarizations accompanying this magnetic evolution are also estimated. Here both the inverse DMI [$\sim \sum_{<ij>} \mathbf{e}_{ij} \times (\mathbf{S}_i \times \mathbf{S}_j)$] [21], and the exchange striction [$\sim \sum_{i,u=x/y} (-1)^i [\mathbf{S}_i \cdot (\mathbf{S}_{i+u} - \mathbf{S}_{i-u})]\mathbf{v}$ ($v = y/x$) [22], are considered for the generation of polarizations [33]. First, the *ab*-plane spiral phase can lead to a ferroelectric polarization along the *a*-axis, generated by the inverse DMI as expected. Second, the intermediate conical phase is nonpolar. Third, the orthogonal stripy phase restores the ferroelectricity along the *a*-axis, although the underlying mechanism is replaced by exchange striction. The polarization finally fades away in the high-field ferromagnetic phase. Typically, the polarization produced by the spiral order is smaller then $\sim 0.1$ μC/cm$^2$, while the exchange striction mechanism dominated polarization can be up to $\sim 1$ μC/cm$^2$ [43,44]. As a reasonable simplification, a weighted parameter $P = p_1 + 10p_3$ is introduced to describe the evolution of polarization plotted in Fig. 4b, indicating both two FE phases probably host quantitatively similar values of polarization. In short, our model simulation indeed reproduces the reentrance of ferroelectricity, i.e. the nontivial polar-nonpolar-polar-nonpolar evolution as well as the possible hidden spin-orthogonal double stripy phase.

## 5. Conclusion

In summary, a hidden type-II multiferroic phase with possible orthogonal spin stripes and large polarization has been revealed in GdMnO$_3$ crystal under high magnetic fields. A full magnetoelectric phase diagram has been established, which is also well reproduced by the model simulation. The spin texture and associated magnetoelectric mechanism are totally different from other type-II multiferroics with reentrance under high field. Our study suggests that unexpected quantum states with multiferroicity may exist even in those well-studied prototype multiferroics.

**Declaration of competing interest**

The authors declare that they have no conflicts of interest in this work.

**CRediT authorship contribution statement**

**Ming Yang:** Writing – review & editing, Writing – original draft, Methodology, Investigation, Formal analysis, Data curation, Conceptualization. **Jun Chen:** Writing – original draft, Visualization, Software. **Junfeng Wang:** Writing – review & editing, Writing – original draft, Resources, Project administration, Methodology, Funding acquisition, Conceptualization. **Chao Dong:** Methodology, Formal analysis, Data curation. **Chengliang Lu:** Methodology, Conceptualization. **Gang Xu:** Writing – original draft, Validation, Software. **Jinguang Cheng:** Writing – original draft, Supervision, Resources, Funding acquisition. **Jianshi Zhou:** Writing – original draft, Supervision, Methodology. **Shuai Dong:** Writing – original draft, Supervision, Software, Project administration, Methodology.

**Acknowledgments**

This work is supported by the National Key Research and Development Plan Project of China (2024YFA1611200), the National Natural Science Foundation of China (12074135, 12004122, 11834002, 12474085, 12025408 and 11874400) and the Fundamental Research Funds for the Central Universities (2019kfyXJJ009). This work is also partly supported by the interdisciplinary program of Wuhan National High Magnetic Field Center (WHMFC202205 and WHMFC202504), Huazhong University of Science and Technology. We thank the Big Data Center of Southeast University for providing the facility support on the numerical calculations.

**Supplementary material**

Supplementary material associated with this article can be found, in the online version, at 10.1016/j.fmre.2025.12.014

## Author profile

**Ming Yang** (BRID: 07227.00.81070) is an associate professor in the Wuhan National High Magnetic Field Center (WHMFC), Huazhong University of Science and Technology. He obtained his PhD from the Laboratoire National des Champs Magnétiques Intenses (LNCMI) in Toulouse, France, and joined WHMFC in 2018. His research interests mainly focus on exploring the magnetic characteristics of different material systems, such as multiferroics, superconductors and charge density wave compounds.

**Junfeng Wang** (BRID: 09668.00.26218) is a professor, PhD supervisor and deputy director of the Wuhan National High Magnetic Field Center (WHMFC), Huazhong University of Science and Technology. He obtained his PhD in 2006 from the Department of Physics, Wuhan University. After completing the postdoctoral work in the international MegaGauss Science Laboratory of the University of Tokyo, he joined the current university. His research interests mainly focus on multiferroics and strongly correlated electrons employing various measurement techniques in pulsed high magnetic fields.

**Shuai Dong** (BRID: 09775.00.89399) is a Chair Professor at Southeast University, where he currently serves as the Chair of the School of Physics. As a theorist, his main research interests focus on quantum materials, especially those exhibiting ferroic properties. He received his PhD degree from Nanjing University and studied/worked as a visiting student, and later as a visiting assistant professor/scientist, at the University of Tennessee at Knoxville and Oak Ridge National Laboratory. He is a recipient of NSFC Distinguished Young Scholars. Prior to joining the Editorial Board of *PRL*, he served as an associate editor for *NPJ Quantum Materials*.